\documentclass[pra,superscriptaddress,floatfix,amsmath,footinbib,amssymb,twocolumn,nofootinbib]{revtex4}
\usepackage{amssymb}
\usepackage{amsmath}
\usepackage{epsfig}
\usepackage{t1enc}
\usepackage{soul}
\usepackage{color}

\usepackage{graphicx}
\usepackage{subfig}
\usepackage{ulem}
\usepackage{caption}

\date{}

\begin{document}

\title{Effect of relativistic motion on witnessing nonclassicality of quantum states}
\author{Agata Ch{\k{e}}ci{\'n}ska}
\email{Agata.Checinska@fuw.edu.pl}
\author{Krzysztof Lorek}
\email{Krzysztof.Lorek@fuw.edu.pl}
\author{Andrzej Dragan} 
\email{Andrzej.Dragan@fuw.edu.pl}
\affiliation{Institute of Theoretical Physics, University of Warsaw, Pasteura 5, 02-093 Warsaw, Poland}

\date{\today}
\begin{abstract}
We show that the operational definition of nonclassicality of a quantum state depends on the motion of the observer. We use the relativistic Unruh-DeWitt detector model to witness nonclassicality of the probed field state. It turns out that the witness based on the properties of the $P$-representation of the quantum state depends on the trajectory of the detector. Inertial and non-inertial motion of the device have qualitatively different impact on the performance of the witness.
\end{abstract}

\maketitle


\section{Introduction }
The behavior of quantum-mechanical systems within relativistic settings is a subject of ongoing research, with the Unruh effect being a prime aspect of these explorations \cite{Unruh-effect,Unruh-effect-1977,Unruh-effect-Fulling,Unruh-effect-review}. Questions raised by relativistic quantum information go beyond the problem of relative particle content and hold a promise of unveiling general properties of quantum mechanics when exposed to non-inertial motions and strong gravitational fields. This includes analysis of detection of entanglement by non-inertial observers \cite{localized}, extraction of entanglement from the vacuum \cite{reznik,extra}, generation of entanglement due to motion \cite{RQ-gates} and improvement of relativistic metrological protocols \cite{RQ-metrology}. It also involves discussion of universal decoherence due to the gravitation dilation \cite{RQ-decoherence}, fundamental limits to building ideal clocks \cite{Ideal-clocks}, experimental simulation closed time-like curves \cite{timelike-curves} and entangling power of the expanding universe \cite{entangling-power}, proposals for verification of the space-time topology \cite{Spacetime-topology} and relativistic protocols within circuit quantum electrodynamics architecture \cite{super}.   

Here we would like to discuss one of the quantum features, namely the nonclassicality of quantum states, and ask how its operationally defined measure is influenced by general motions. nonclassicality of quantum states can be linked to a variety of phenomena observed experimentally \cite{exp1,exp2,exp3, wahadlo}, such as photon antibunching \cite{antibunching}, sub-Poissonian photon statistics \cite{sub-Poissonian}, squeezing \cite{squeezing}, and can be considered a resource \cite{wilde}. There is a number of different approaches to this concept \cite{P-classicality,approaches-nonclassicality,Vogel,Diosi,Richter}. However, no unique definition exists. 
We are motivated by the approach taken in the field of quantum optics where the criterion for nonclassicality is frequently based on the properties of the $P$-representation. What we are interested in, is verifying whether such a criterion, defined operationally, is sensitive to the motion of the probe detecting nonclassicality. We show that indeed such sensitivity occurs and the universality of the operational criterion is further limited by a presence of relative motions. Even the Minkowski vacuum state, which is considered classical from the perspective of an inertial frame, can yield non-classical properties when viewed by an accelerated observer, associated with a different basis \cite{reznik}. The purpose of this work is to show explicitly that operational measures of nonclassicality such as the one studied in this paper, are actually observer-dependent.
In order to study such effects let us first review the necessary formalism behind the $P$-representation. Any state of a harmonic oscillator can be represented in the basis of coherent states, the so-called diagonal basis:
\begin{align}  
\rho_{\text{ho}}&=\int \text{d}^2\alpha P(\alpha)|\alpha\rangle\langle\alpha|,
\end{align}
where $P(\alpha)$ is the Glauber-Sudarshan diagonal $P$-representation, which is a probability quasi-distribution function \cite{Sudarshan, Glauber}.
Coherent state represents the closest approximation of a
classical oscillator, with the minimum uncertainty and oscillating expectation value of the position and the momentum \cite{coherent-states}.
Therefore a quantum state $\rho_{\text{ho}}$, which can be represented as a statistical mixture of coherent states, can be considered classical \cite{P-classicality}. This occurs when the corresponding $P$-representation $P(\alpha)$ satisfies the properties of a probability distribution function and is a positive definite function. However, when $P(\alpha)$ takes on negative values or is highly singular, this statement does not hold anymore. 

It is possible to derive the $P$-representation from the full quantum state tomography \cite{P-tomography}. However, in general this procedure is highly sensitive to experimental errors and thus not practical. For that reason, the solution is to search for measures that can tune in to the properties of $P$-representation without the necessity of using state tomography.
In this body of work we are interested in the quantum-optical approach that leads to an operationally defined witness of nonclassicality. We follow the proposal presented in \cite{Eberly} where the authors introduced a witness based on a two-level probe interacting with a harmonic oscillator. Similarly to that we probe a state of a quantum field with a two-level detector and infer the presence of the aforementioned nonclassicality of this state from the detector's readings. Our analysis reveals how different motions of the detector affect  the performance of the witness based on the $P$-representation.

The paper is structured as follows: in Sec.~\ref{s2} we recall the formalism of the witness of nonclassicality, in Sec.~\ref{s3} we discuss the impact of relative motion of the detector and the measured system on the performance of the witness and in Sec.~\ref{s4} we give a summary and an outlook. Throughout the paper we use natural units with $c=\hbar=1$.

\section{Witness of nonclassicality\label{s2}}

Let us start by briefly reviewing the operational witness of nonclassicality based on the approach presented in \cite{Eberly}. We are interested in the operational evaluation of nonclassicality of an unknown initial state of a quantum system. We consider a qubit detector which interacts with a harmonic oscillator. After the interaction, by analyzing the state of the detector we are able to verify the presence of nonclassicality in the initial state of the harmonic oscillator.\\
The system is assumed to start in a product state $\rho(0)=\rho_{\text{qb}}(0)\otimes\rho_{\text{ho}}(0)$, corresponding to the qubit and the harmonic oscillator, and evolve according to a unitary transformation $\hat{U}=\exp\{-it\hat{H}\}$ generated by a Hamiltonian: 
\begin{align}
\label{introham}
\hat{H}&=\omega_{\text{qb}}\hat{\sigma}_z+\omega_{\text{ho}}\hat{a}^\dagger \hat{a}+ \lambda \hat{\sigma}_z(\hat{a}+\hat{a}^\dagger).
\end{align} 
In the above equation $\hat{a},\,\hat{a}^\dagger$ denote the usual bosonic annihilation and creation operators, $\hat{\sigma}_i$ are the Pauli operators, $\omega_{\text{qb}}$, $\omega_{\text{ho}}$ are the frequencies of the respective subsystems, and $\lambda$ is the coupling constant between them.
We choose to parametrize the initial state in the following manner:
\begin{align}
\rho(0)&=\left(\begin{array}{cc}z(0)&w(0)\\w^*(0)&1-z(0)\end{array}\right)\otimes \int \text{d}^2\alpha P(\alpha) |\alpha\rangle\langle\alpha|\nonumber\\
&\equiv\rho_{\text{qb}}(0)\otimes\rho_{\text{ho}}(0),
\end{align}
where we have used the diagonal representation for the harmonic oscillator's density operator $\rho_{\text{ho}}(0)$ and the eigenbasis of $\sigma_z$ for the qubit's operator $\rho_{\text{qb}}(0)$. The full density operator evolves according to the equation $\rho(t)=e^{-i\hat{H}t}\rho(0)e^{i\hat{H}t}$ and the state of the detector at any time $t$ is given by the reduced density operator $\rho_{\text{qb}}=\text{Tr}_{\text{ho}}\rho(t)$. Since $[\hat{\sigma}_z,\hat{H}]=0$, the only non-trivially evolving elements of the qubit density matrix, $\rho_{\text{qb}}(t)$, are the off-diagonal ones: 
\begin{align}
w(t)=\text{Tr}\{|0\rangle\langle 1|\rho(t)\},
\end{align}
where $|1\rangle,\,|0\rangle$ denote $\hat{\sigma}_z$ eigenstates with corresponding eigenvalues $\{\pm 1\}$. An explicit calculation of the evolution of $w(t)$ yields the following result:
\begin{align}
|w(t)|&=|w(0)|e^{-8\left(\lambda/\omega_{\text{ho}}\right)^2\sin^2(\omega_{\text{ho}}t/2)}|W(t)|,
\end{align}
where the function:
\begin{align}
W(t)&\equiv \int \text{d}^2\alpha P(\alpha)e^{-4i\frac{\lambda}{\omega_{\text{ho}}}(\alpha e^{-i\omega_{\text{ho}} t/2}\,+\,\alpha^*e^{i\omega_{\text{ho}} t/2})\sin\frac{\omega_{\text{ho}} t}{2}},\label{witness_def}
\end{align}
is the \textit{witness function}\footnote{For brevity, we may also refer to $|W(t)|$ as the witness function.}. If the underlying $P$-representation is positive, then one can write the \textit{witness inequality} (or the \textit{classicality bound}): 
\begin{align}
|W(t)|&= e^{8\left(\lambda/\omega_{\text{ho}}\right)^2\sin^2(\omega_{\text{ho}}t/2)}\left|\frac{w(t)}{w(0)}\right|\leq 1, \label{mono-witness}
\end{align}
which does not depend on the choice of $\omega_{\text{qb}}$ and without loss of generality we can consider gapless detectors\footnote{The choice of a gapless detector has the advantage of simplifying calculations; introducing a small energy gap would require applying a perturbative approach to solve the evolution of the system.} with $\omega_{\text{qb}}=0$.
The upper bound on the measurable quantity $|W(t)|$ is therefore related to the characteristics of the $P$--representation that describes the initial state of the harmonic oscillator. Observed violation of the witness inequality indicates that the $P$-representation of the initial state $\rho_{\text{ho}}(0)$ has been negative, thus non-classical. However, a lack of this violation does not guarantee that $\rho_{\text{ho}}(0)$ can be considered classical, as shown in particular examples in \cite{Eberly}.

For a better understanding of the witness function let us look at quantum states that reveal their nonclassicality when tested within this formalism. 
This category includes Fock states and Schr\"{o}dinger Cat states. The $P$-representation corresponding to a Fock state $|N\rangle$ is given by:
\begin{align}
P_N(\alpha)&=\frac{e^{|\alpha|^2}}{N!}\bigg(\frac{\partial^{2N}}{\partial\alpha^N\partial\alpha^{*N}}\delta^2(\alpha)\bigg),\label{P-Fock}
\end{align}
and becomes singular for $N>0$, meaning that out of all Fock states, only the vacuum has a classical $P$-representation. The witness function $W(t)$ evaluated for the Fock state $|N\rangle$ can be neatly expressed via Laguerre polynomials $L_N(x)$ as:  
\begin{align}
W_N(t)&=L_N\bigg(16\frac{\lambda^2}{\omega_{\text{ho}}^2}\sin^2\frac{\omega_{\text{ho}} t}{2}\bigg).\label{witness-fock-N}
\end{align}
The second example is provided by the Schr\"{o}dinger Cat states, which are non-classical despite being an equal superposition of two coherent states which individually  are considered classical:
  \begin{align}
|\psi_{\text{SC}}(\alpha_0)\rangle&=\,\frac{|\alpha_0\rangle+|-\alpha_0\rangle}{\sqrt{2(1+e^{-2\alpha_0^2})}}.
 \end{align}
The corresponding $P$-representation is given by: 
 \begin{align}
P_{\text{SC}}(\alpha)&=\bar{N}^2\big[\delta^2(\alpha-\alpha_0)+\delta^2(\alpha+\alpha_0)+e^{|\alpha|^2-|\alpha_0|^2}\nonumber\\
&\times\big(e^{\alpha_0\partial_{\alpha^*}}e^{-\alpha_0\partial_\alpha}+e^{-\alpha_0\partial_{\alpha^*}}e^{\alpha_0\partial_\alpha}\big)\delta^2(\alpha)\big],\label{P-SC}
\end{align}
 with $1/\bar{N}^2=2(1+e^{-2\alpha_0^2})$. 
\begin{center}
\begin{figure}
\includegraphics{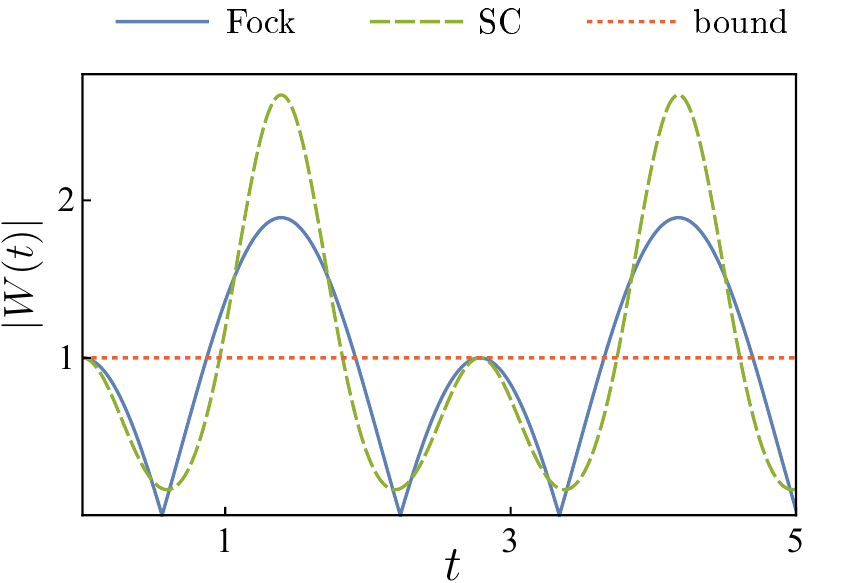}
\caption{\label{fig1} Witness function $|W(t)|$ for the Fock state $|1\rangle$ and the Schr\"odingier Cat state $|\psi_{\text{SC}}(\alpha_0)\rangle$, and the following set of parameter values: $\lambda=1.7$, $\omega_{\text{ho}}=4/\sqrt{\pi}$, $m=1$, $L=2$ and $\alpha_0=1$.  }
\end{figure}
\end{center}
The witness functions for both a Fock state and a Schr\"{o}dinger Cat state are plotted as a function of time in Fig.~\ref{fig1}. The upper bound dictated by the classicality of the $P$-representation is violated periodically.
Moreover, this violation, if present, occurs only after a certain minimum time and for $\lambda$ above a certain threshold value \cite{Eberly}. \\


\section{nonclassicality and relative motion\label{s3}}
In this section we analyze how the witness of nonclassicality, as introduced in the previous section, is affected if the detector and the quantum system of interest are in an inertial or accelerated motion.  
To probe the properties of the state of a quantum field we employ a model of a detector used in quantum field theory on a curved space-time, the so-called Unruh-DeWitt detector \cite{Unruh-effect,DeWitt}. This model involves a point-like, semi-classical detector with two internal energy levels (qubit), following a classical trajectory $x(\tau)$, where $\tau$ is the detector's proper time \cite{bd}. In our scenario we probe a real, 1-D, scalar quantum field $\hat{\phi}$ (massive or massless), satisfying the Klein-Gordon equation and confined in a resting cavity of length $L$. We set the left wall of the cavity at $x=0$ and the initial position of the detector at $x(\tau=0)=x_0=\frac{L}{2k}$, which is at the leftmost antinode of the field. This choice will be explained later. The Hamiltonian of the Unruh-DeWitt detector interacting with the field is given by:
\begin{align}
\hat{H}_I(\tau)&=\lambda\epsilon(\tau)\hat{\phi}[x(\tau)](\hat{\sigma}_+e^{i\omega_{\text{qb}}\tau}+\hat{\sigma}_-e^{-i\omega_\text{qb}\tau}),
\end{align}
where $\hat{\sigma}_\pm$ denote the detector's rising and lowering operators, $\lambda$ is the coupling strength, $\epsilon(\tau)$ is a smooth switching function, that we assume to be $\epsilon(\tau)\approx 1$ during the interaction and decreasing to zero when $\tau\rightarrow\pm\infty$. Furthermore, $\hat{\phi}[x(\tau)]$ is the field operator of the quantum field probed by the detector along the classical trajectory $x(\tau)$, parametrized by the detector's proper time $\tau$. We are interested in the scenario wherein the cavity is resting while the detector is in motion, and we expand the field operator in terms of the solutions of the Klein-Gordon equation written in the inertial reference frame of the cavity \cite{Accelerometer,Regula2016} (and in the interaction picture):
\begin{align}
\hat{\phi}[x(\tau)]&=\sum_kF_k(x(\tau))(\hat{a}_ke^{-i\omega_k \tau}+\hat{a}_k^\dagger e^{i\omega_k \tau}),
\end{align}
where the mode functions are given by:
\begin{align}
F_k(x(\tau))&=\frac{1}{\sqrt{k\pi}}\sin\left(\frac{k\pi}{L}x(\tau)\right),
\end{align}
and the frequencies are defined as $\omega_k=\sqrt{(k\pi /L)^2+m^2}$, with $m$ being the mass of the field. In the above, we have chosen a standard basis for the decomposition of the field operator, i.e. plane waves. However, this is not the only basis allowed.

For the gapless Unruh-DeWitt detector model, $\omega_{\text{qb}}=0$, the Hamiltonian takes a much simpler form. In the Schr\"{o}dinger picture the total Hamiltonian of the system can be written as a sum $\hat{H}(\tau)=\sum_k\hat{H}_k(\tau)$ with:
\begin{align}
\hat{H}_k(\tau)=\omega_k \hat{a}^\dagger_k\hat{a}_k+\lambda F_k(x(\tau))(\hat{a}_k+\hat{a}_k^\dagger)\hat{\sigma}_x.\label{H}
\end{align}
This allows us to write the evolution operator in the following way:
\begin{align}
\hat{U}(\tau)&=\bigotimes_k\hat{U}_{k}(\tau),\qquad\hat{U}_{k}(\tau)=\mathcal{T}e^{-i\int_{0}^\tau \text{d}\tau' \hat{H}_{k}(\tau')}.\label{U-k}
\end{align}
We assume that the system starts in a separable state that can be written in the form:
\begin{align} 
\rho(0)&=\rho_{\text{qb}}(0)\otimes\rho_\phi(0)&=\left(\begin{array}{cc} p(0) & w(0)\\ w^*(0) & 1-p(0)\end{array}\right)\otimes\rho_\phi(0),
\end{align}
where the qubit's density operator, $\rho_{\text{qb}}(0)$, is given in the eigenbasis of $\hat{\sigma}_x$, $|\pm\rangle$, with its corresponding eigenvalues equal to $\pm 1$, and $\rho_\phi(0)$ denotes the density operator of the field. The full state of the system evolves according to:
\begin{align}
\rho(\tau)=\hat{U}(\tau)\rho(0)\hat{U}^\dagger(\tau).
\end{align} 

The time evolution generated by our Hamiltonian is the well-known evolution of a forced harmonic oscillator \cite{fho}. Based on this we can obtain the explicit form of the transformation $\hat{U}_k(\tau)$ (see Appendix A for further details): 
\begin{align}
\hat{U}_k(\tau)&=\,e^{i\gamma_k(\tau)}\hat{D}\left(\chi_k(\tau)e^{-i\omega_k\tau}\right)e^{-i\omega_k\tau \hat{a}^\dagger_k \hat{a}_k},
\end{align}
where:
\begin{align}
\gamma_k(\tau)&=\,\lambda^2 \int_0^\tau\int_0^{\tau'}\text{d}\tau'\text{d}\tau''F_k(x(\tau'))F_k(x(\tau''))\times\nonumber\\
&\qquad\qquad\qquad\times\sin(\tau'-\tau''),\\
\chi_k(\tau)&=-i\lambda\int_0^\tau d\tau'F_k(x(\tau))e^{i\omega_k\tau'},\label{chi}
\end{align}
and $\hat{D}(\alpha)$ is the displacement operator $\hat{D}(\alpha)=\exp\{\alpha \hat{a}^\dagger-\alpha^*\hat{a}\}$.

We follow the same steps as in the previous section and examine the evolution of the detector's density operator: $\rho_{\text{qb}}(\tau)=\text{Tr}_\phi\rho(\tau)$. Our goal is to study how non-trivial trajectories of the detector affect the performance of the nonclassicality witness introduced in the previous section. Note that for the special case of the detector at rest the Unruh-DeWitt detector model reduces to the previously studied Hamiltonian \eqref{introham}.

Again, the only non-trivially evolving elements of $\rho_{\text{qb}}(\tau)$ are the off-diagonal elements $w(\tau)$ (since $[\hat{\sigma}_x,\hat{H}(\tau)]=0$): 
\begin{align}
w(\tau)&=\,\text{Tr}\big\{|-\rangle\langle +|\otimes\mathbb{I}\,\rho(\tau)\big\}\nonumber\\
&=
\text{Tr}\big\{\hat{U}^\dagger(\tau)|-\rangle\langle +|\hat{U}(\tau)\rho_{\text{qb}}(0)\otimes\rho_{\phi}(0)\big\}\label{w-mid}\\
&=\text{Tr}_{\text{qb}}\big\{|-\rangle\langle +|\rho_{\text{qb}}(0)\big\}\text{Tr}_\phi\big\{\hat{U}_+(\tau)\rho_\phi(0)\hat{U}_-^\dagger(\tau)\big\},\nonumber
\end{align}
where $\hat{U}_\pm$ is the time evolution operator $\hat{U}(\tau)$, in which the operator $\hat{\sigma}_x$ is replaced with the corresponding eigenvalues $\hat{\sigma}_x\to \pm 1$. Since $\text{Tr}_{\text{qb}}\{|-\rangle\langle +|\rho_{qb}(0)\}=w(0)$, we can write \eqref{w-mid} as:
\begin{align}
w(\tau)&=w(0)\text{Tr}_\phi\big\{\hat{U}_+(\tau)\rho_\phi(0)\hat{U}_-^\dagger(\tau)\big\},
\end{align}
and 
\begin{align}
\frac{w(\tau)}{w(0)}&=\text{Tr}_\phi\big\{\hat{U}_+(\tau)\rho_\phi(0)\hat{U}_-^\dagger(\tau)\big\}.
\end{align}
This gives us the ratio appearing in Eq.~\eqref{mono-witness}.

To see how the witness performs when relative motion is introduced, we initially prepare the field in certain test states. 
We choose those states such that they are regarded non-classical by a resting detector. In particular, we analyze Fock states and Schr\"{o}dinger Cat states occupying the $k$-th mode of the cavity and assume that all the remaining modes are in the vacuum state. Therefore we write: 
\begin{align}
\rho_\phi(0)&=\bigotimes_k\int \text{d}^2\alpha P_k(\alpha)|\alpha\rangle\langle \alpha|,
\end{align}
where $P_{k\neq k_0}(\alpha)=\delta(\alpha)$ and $P_{k_0}(\alpha)$ is the $P$-representation of either Fock state or Schr\"{o}dinger Cat state. With such a choice of the initial state we write:
\begin{align}
\left|\frac{w(\tau)}{w(0)}\right|&=\prod_k|\int \text{d}^2\alpha P_k(\alpha)\text{Tr}_k\big\{\hat{U}_{k,+}(\tau)|\alpha\rangle\langle\alpha|\hat{U}_{k,-}^\dagger(\tau)\big\}|\nonumber\\
&=\prod_k|\int \text{d}^2\alpha P_k(\alpha)e^{4i\operatorname{Im}(\alpha^*\chi_k(\tau))-2|\chi_k(\tau)|^2}|\nonumber\\
&=e^{-2\sum_k|\chi_k(\tau)|^2}|\int \text{d}^2\alpha P_{k_0}(\alpha)e^{4i\operatorname{Im}(\alpha^*\chi_{k_0}(\tau))}|.
\end{align}
This equation explicitly shows the relation between the qubit readings and the $P$-representation of the initial test state. 
Next, we introduce the witness function characterizing our test state:
\begin{align}
W_{k_0}(\tau)&\equiv\int d^2\alpha P_{k_0}(\alpha)e^{4i\text{Im}(\alpha^*\chi_{k_0}(\tau))} \nonumber\\
&=\frac{w(\tau)}{w(0)}e^{2\sum_k|\chi_k(\tau)|^2}.\label{witness-moving}
\end{align}
Since for classical states we have $|P_{k_0}(\alpha)|=P_{k_0}(\alpha)$, we can evaluate an upper bound for the absolute value of the witness function:
\begin{align}
|W_{k_0}(\tau)|&=\left|\int d^2\alpha P_{k_0}(\alpha)e^{4i\text{Im}(\alpha^*\chi_{k_0}(\tau))}\right|\nonumber\\
& \leq \int d^2\alpha |P_{k_0}(\alpha)|\left|e^{4i\text{Im}(\alpha^*\chi_{k_0}(\tau))}\right| \nonumber\\
&=\int d^2\alpha |P_{k_0}(\alpha)|=1.
\end{align}
A violation of this inequality indicates that the initial state of the field has been non-classical due to $|P_{k_0}(\alpha)|\neq P_{k_0}(\alpha)$. 
Using this, we analyze how the trajectory of the detector influences the witness function. 

We start with examining two examples of states which are considered classical by a resting detector, i.e. a coherent and a thermal state. A simple calculation shows that these are also identified as classical from the perspective of a moving detector, inertial or non-inertial (see Appendix B for the details). Thus for these states the performance of the witness stays unaffected by the motion of the detector.

\paragraph{Fock states.} Let us now investigate a Fock state with a single excitation, $|0\rangle_{k\neq k_0}\otimes|1\rangle_{k_0}$. This state of the field is described by the $P$-representation given in Eq.~\eqref{P-Fock}. We look at two scenarios: I) the detector moves with a constant velocity $v$ until it hits the right wall of the cavity; its trajectory is given by $x_v(\tau)=\frac{v\tau}{\sqrt{1-v^2}}+\frac{L}{2k_0}$;\\ II) the detector moves with a constant acceleration $a$ until it stops at the right wall; its trajectory is described by $x_a(\tau)=\frac{1}{a}[\cosh(a\tau)-1]+\frac{L}{2k_0}$. In both cases the detector starts at the leftmost antinode of the field, $x_0=\frac{L}{2k_0}$. This choice is made since we wish to maximize the coupling between the detector and the field and the duration of their interaction.

For both types of trajectories we obtain the witness function of the form similar to Eq.~\eqref{witness-fock-N}, which for one-particle states simplifies to:  
\begin{align}
 W_{k_0}^F(\tau)&=1-4|\chi_{k_0}(\tau)|^2,
 \end{align}
with $\chi_{k_0}(\tau)$ defined in Eq.~\eqref{chi}. A closer inspection of $\chi_{k_0}(\tau)$ shows that its dependence on parameters $k_0, L, \lambda$ enters only through $\frac{k_0}{L}$ and $\frac{\lambda}{\sqrt{k_0}}$. Due to this property, increasing $k_0$ while keeping $\lambda\sim\sqrt{k_0}$ and $L\sim k_0$, leaves the witness function unchanged but allows for an increased time of interaction. With no loss of generality we choose to work in the regime of large $L$ and $k_0$. Moreover, we take a value of $\lambda$ that guarantees violation of the classicality bound for a corresponding non-moving detector, unless stated otherwise.

Let us look at the first scenario. It is possible to derive an analytical formula for $W_{k_0}(\tau)$ revealing the periodicity of the witness function, due to a fairly simply form of $\chi_{k_0}(\tau)$:
\begin{align}
\chi_{k_0}(\tau)=\frac{-i\lambda}{\sqrt{k_0\pi}}\int_0^\tau &\text{d}\tau'\sin\left(k_0\pi\left(\frac{1}{2}+\frac{x_0}{L}\right)+\frac{k_0\pi v\,\tau'}{L\sqrt{1-v^2}}\right)\nonumber\\
&\times e^{i\omega_{k_0}\tau'}.\label{chi_v}
\end{align}
The resulting expression is fully discussed in Appendix C. In Fig.~\ref{Fock-v} we plot $W_{k_0}(\tau)$ evaluated for an initial one-particle Fock state for a detector moving with different velocities $v$. The behavior of the witness function remains oscillatory throughout the entire dynamics and thus allowing for a periodic violation of the classicality bound. In the formula above we see two typical frequencies: $\omega_{k_0}$ and the frequency with which the detector passes through the consecutive maxima of the mode function, $\omega_L=\frac{k_0\pi v}{L\sqrt{1-v^2}}$. When these two match, a resonance occurs. This is shown in the plot of the time-averaged witness function $\bar{W}_v=\frac{1}{T_1-T_2}\int_{T_1}^{T_2} d\tau |W_{k_0}(\tau)|$ against the velocity of the detector, in Fig.~\ref{Fock-vs-v}. Equating the two frequencies yields a critical velocity for a given wavenumber $k_0$:
   \begin{figure}
   \includegraphics{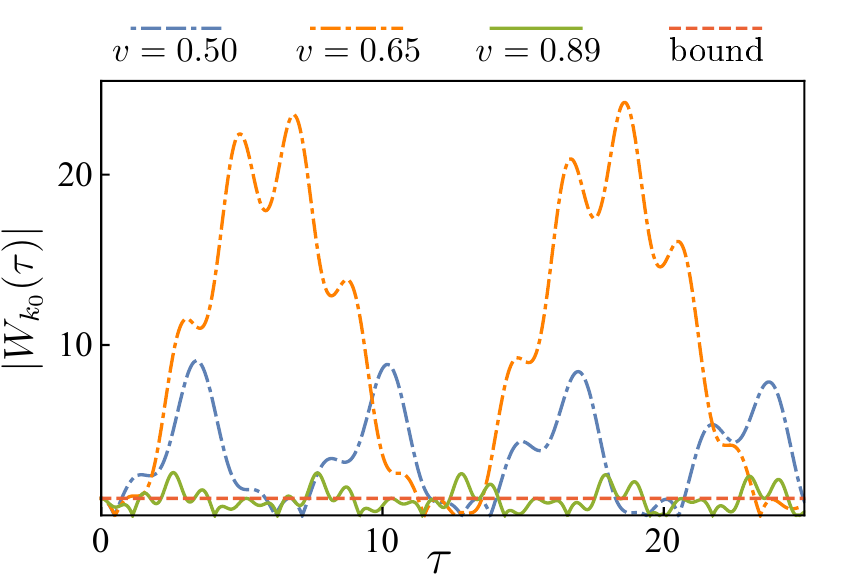}
 \caption{\label{Fock-v} Witness function $|W_{k_0}(\tau)|$ corresponding to the initial Fock state $|\Psi\rangle_{k_0}=|0\rangle_{k\neq k_0}\otimes |1\rangle_{k_0}$ for the detector moving with a constant velocity $v$ and the following set of parameters: $k_0=5000$, $L=10000$, $x_0=L/2k_0$, $\lambda=2\sqrt{k_0}$ and $m=1$. }
 \end{figure}
 \begin{align}
 v_{c}=\sqrt{\frac{1+(k_0\pi/m L)^2}{1+2(k_0\pi/mL)^2}}.\label{v-c}
 \end{align}
 This resonant behavior persists also for small couplings $\lambda$ for which a resting detector does not violate the classicality bound.  When the critical velocity is approached we observe a simultaneous increase in the period of oscillations of the witness function. We suspect that the occurrence of this resonance is due to the specific method used to investigate a nonclassicality measure and might not emerge in other scenarios. Furthermore, we note that the performance of this witness is enhanced, meaning that even when nonclassicality is not detected with a stationary detector, it might be seen by a moving one. Relativistic velocities are needed, however, for this effect to be observable.
  \begin{figure}
  \includegraphics{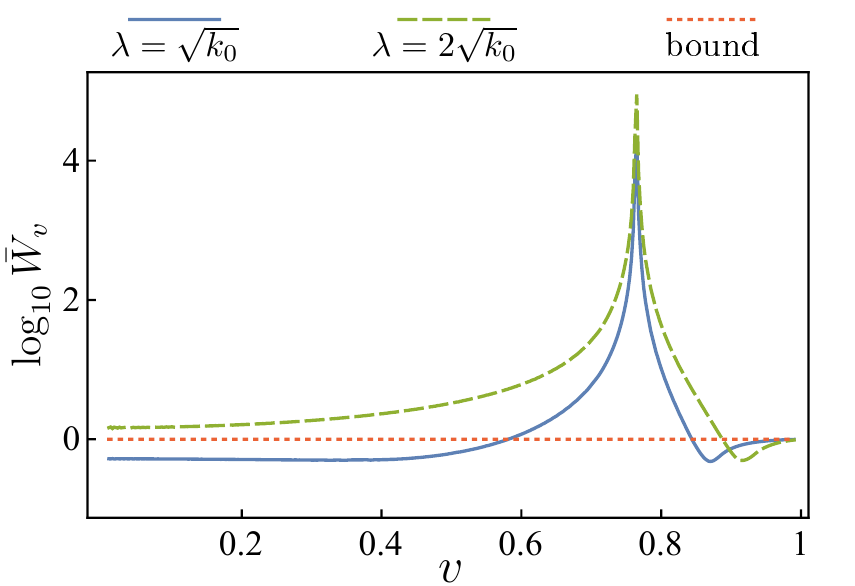}
 \caption{\label{Fock-vs-v} Time-averaged $|W_{k_0}(\tau)|$ as a function of velocity $v$, with averaging time window $\tau\in\{0,500\}$ for the following parameter values: $k_0=5000$, $L=10000$, $x_0=L/2k_0$, $\lambda=2\sqrt{k_0}$ and $m=1$. The plot is shown in a logarithmic scale. }
 \end{figure}

\begin{figure}
 \includegraphics{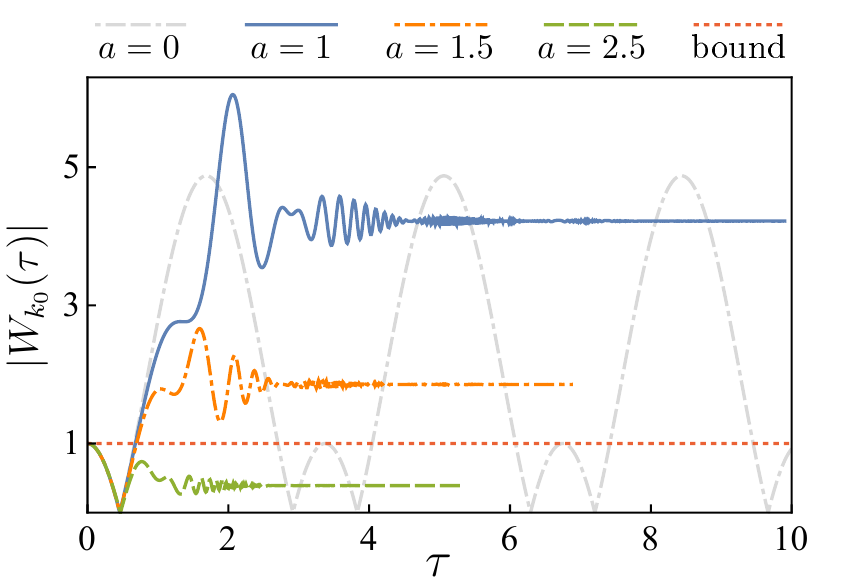}
 \caption{\label{Fock-a} Witness function $|W_{k_0}(\tau)|$ corresponding to the initial Fock state $|\Psi\rangle_{k_0}= |0\rangle_{k\neq k_0}\otimes |1\rangle_{k_0}$ for the detector moving with a constant acceleration, and the following set of parameter values: $k_0=5000$, $L=10000$, $x_0=1$, $\lambda=2\sqrt{k_0}$ and $m=1$.}
 \end{figure}

\begin{figure}
 \includegraphics{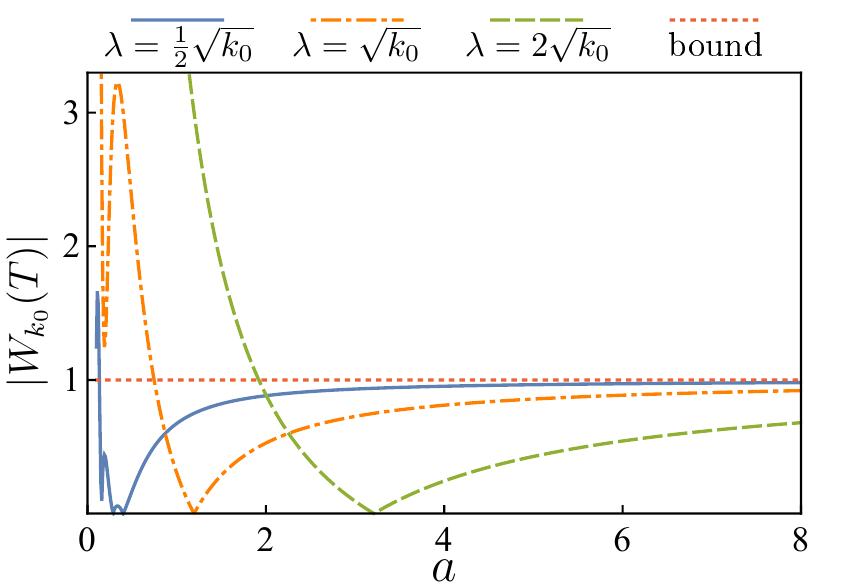}
 \caption{\label{Fock-vs-a} Witness function $|W_{k_0}(T)|$ corresponding to the initial Fock state $|\Psi\rangle_{k_0}=|0\rangle_{k\neq k_0}\otimes |1\rangle_{k_0}$, as a function of acceleration, evaluated at $T=500$ and for the following parameter values: $k_0=5000$, $L=10000$ and $m=1$.}
 \end{figure}

Next, we proceed with the second scenario in which the detector accelerates. In Fig.~\ref{Fock-a} we plot the absolute value of the witness function $|W_{k_0}(\tau)|$ for various accelerations. The behavior is now qualitatively different, as the curves initially oscillate, but after a certain time they start asymptotically approaching constant values. For the regime of parameters investigated, we see that the higher the acceleration of the detector, the earlier the asymptote emerges. Fig.~\ref{Fock-vs-a} shows the dependence of this asymptotic value on the acceleration $a$ for a number of coupling strengths. The violation of the classicality bound is observed only for small accelerations, for which the value of the asymptote initially oscillates with $a$. For higher accelerations the value of the asymptote tends to $1$ from below. Thus small accelerations will enhance the performance of the witness, but higher ones will inevitably lead to a situation in which we cannot see the nonclassicality. \\

Finally, it should be noted that discussing small vs. large accelerations requires specifying the relative scale. This is, however, impossible without a broader analytical study of the mathematical properties of the witness function $W(a)$, which we have investigated only numerically. This critical acceleration is related to the point of intersection of the witness function $W(a)$, with the classicality bound. We suspect that this value is connected to the interplay of the expected thermalization (due to the Unruh effect) and the degree to which the state of the detector is affected by the coupling. The witness function approaches the asymptotic behavior after a certain time, and the higher the temperature of the surrounding thermal bath is, the more the detector gets affected. This eventually leads to a classical signature in the witness function. The full understanding of this phenomenon goes beyond the abilities of this simplified model and requires further studies.


 %
 %
\begin{figure}
\includegraphics{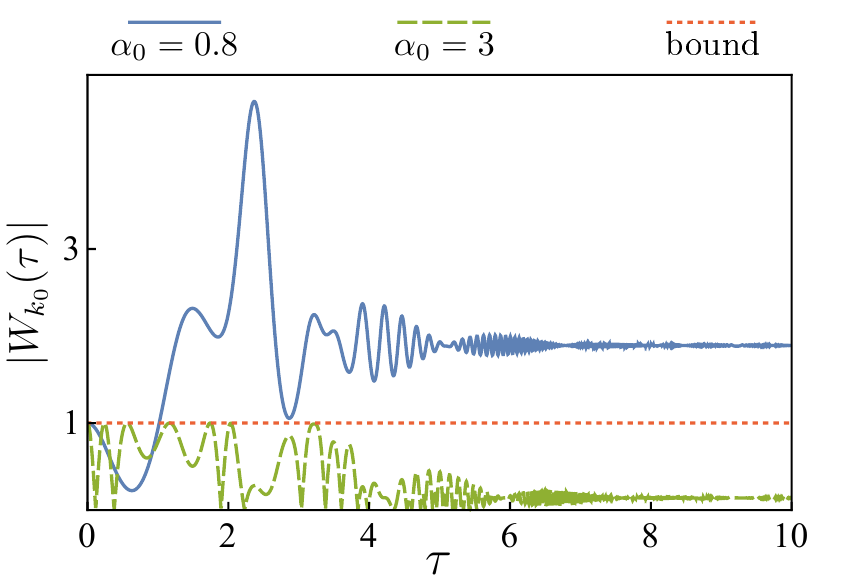}
\caption{\label{sc-al} Witness function $|W_{k_0}(\tau)|$ corresponding to the initial Schr\"odinger Cat state $|\Psi\rangle_{k_0}=|0\rangle_{k\neq k_0}\otimes|\Psi_{\text{SC}}(\alpha_0)\rangle_{k_0}$ for the detector moving with a constant acceleration, and the following set of parameter values: $a=0.8$, $k=5000$, $L=10000$, $x_0=1$, $\lambda=2\sqrt{k_0}$ and $m=1$.}
\end{figure}

\begin{figure}
\includegraphics{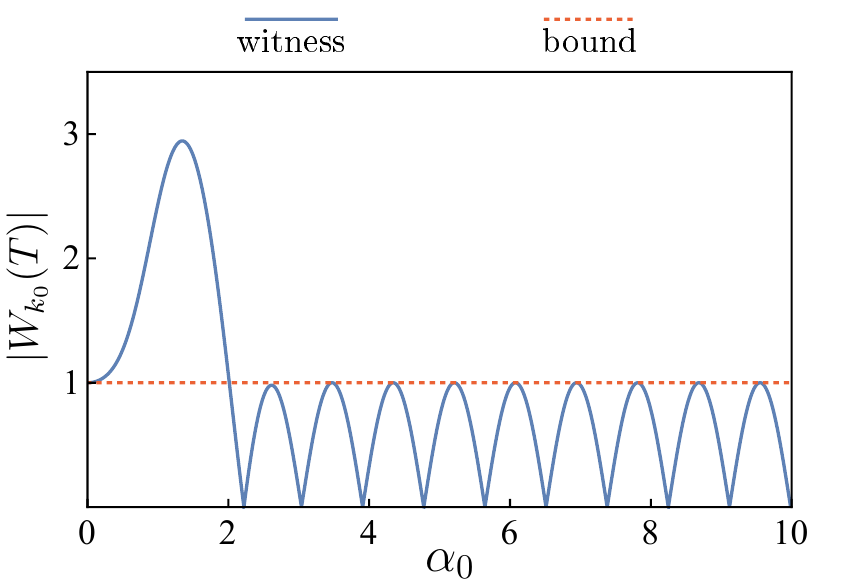}
\caption{\label{sc-vs-al} Witness function $|W_{k_0}(T)|$ corresponding to the initial Schr\"odinger Cat state $|\Psi\rangle_{k_0}=|0\rangle_{k\neq k_0}\otimes|\Psi_{\text{SC}}(\alpha_0)\rangle_{k_0}$ as a function of $\alpha_0$, evaluated at $T=100$ and for the following parameter values: $a=0.8$, $k=5000$, $L=10000$, $x_0=1$, $\lambda=2\sqrt{k_0}$ and $m=1$.}
\end{figure}

\begin{figure}
\includegraphics{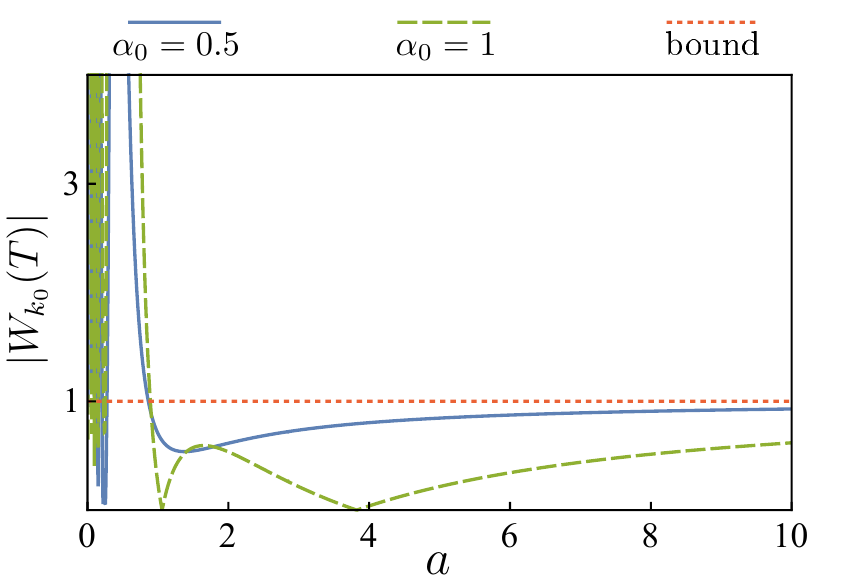}
\caption{\label{sc-vs-a} Witness function $|W_{k_0}(T)|$ corresponding to the initial Schr\"odinger Cat state $|\Psi\rangle_{k_0}=|0\rangle_{k\neq k_0}\otimes|\Psi_{\text{SC}}(\alpha_0)\rangle_{k_0}$ as a function of acceleration, evaluated at $T=1000$ and for the following parameter values: $k=5000$, $L=10000$, $x_0=1$, $\lambda=2\sqrt{k_0}$ and $m=1$.}
\end{figure}

 \paragraph{Schr\"odinger Cat states.} In this paragraph we analyze the second family of quantum states and initiate the field in a Schr\"{o}dinger Cat state, namely $|0\rangle_{k\neq k_0}\otimes|\Psi_{\text{SC}}(\alpha_0)\rangle_{k_0}$. The corresponding $P$-representation is given in Eq.~\eqref{P-SC} and leads to the following form of the witness function:
 \begin{align}
 W_{k_0}^{\text{SC}}(\tau)=&
 \frac{1}{1+e^{-2\alpha_0^2}}  
 \{
 \cos\left[4\alpha_0\text{Im}\chi_{k_0}(\tau)\right] \\&+
 e^{-2\alpha_0^2}\cosh\left[4\alpha_0\text{Re}\chi_{k_0}(\tau)\right]
 \}.\nonumber
 \end{align}
 The numerical analysis has been repeated for the same two trajectories $x_v(\tau)$ and $x_a(\tau)$ to reveal qualitatively the same results. For the inertial motion we observe a resonant behavior similar to the previous case, also characterized by the critical velocity given in Eq.~\eqref{v-c}, since the same $\chi_{k_0}(\tau)$ appears in the formula. For the accelerated motion we see the emergence of the asymptotic behavior for long interaction times, and the behaviour is qualitatively the same as for the Fock case. This can be seen in the Fig.~\ref{sc-al}, which shows the absolute value of the witness function for two values of the parameter $\alpha_0$. Fig.~\ref{sc-vs-al} and Fig.~\ref{sc-vs-a} show in detail the value of the asymptote against parameter $\alpha_0$ and the acceleration $a$ respectively. Similarly to what we have seen before, the value of the asymptote oscillates with $a$ for low accelerations. These oscillations have a high amplitude and are more rapid than in the Fock case, as we observe in Fig.~\ref{sc-vs-a} for $a$'s approaching zero. For large $a$'s the asymptote tends to unity from below. One can see that nonclassicality can be detected only if the acceleration $a$ and the parameter $\alpha_0$ are small. It should be noted that again, the higher the acceleration, the earlier these asymptotes appear. Yet, we have confirmed that the rapid oscillations that we see in the Fig.~\ref{sc-vs-a} as the curves approach zero, are the actual values of the asymptote, i.e. the time $T$, at which the value of the asymptote is evaluated, has been chosen to be sufficiently high.
 

\section{Summary\label{s4}}
In this work we have discussed the influence of inertial and non-inertial motions on the performance of the operationally defined witness of nonclassicality, based on the $P$-representation. Using the model of the Unruh-DeWitt detector we analyzed the performance of the witness for the following test states: Fock and Schr\"odinger Cat states. We have observed a qualitative difference between the inertial and non-inertial case, leading to a conclusion that large accelerations inevitably deteriorate the detection of nonclassicality, whereas inertial motion does not exhibit this property. Our observations also show that applying small accelerations may enhance the detection of nonclassicality, provided that we work with states with high wavenumbers, or equivalently with long cavities. Otherwise our detector hits the cavity wall, before it can positively verify the presence of nonclassicality. The performance of the witness can be enhanced also by inertial motions, even for small coupling strength between the detector and the measured field, due to the resonance effect. This is, however, expected only for velocities comparable to $c$.
Thus we have seen that relative motions can either improve or worsen the detection of nonclassicality. Observation of nonclassicality becomes therefore dependent on the motional state of the device that performs the measurement. 

A number of interesting open questions remains for further studies. One challenge is to better understand how the $P$-representation transforms under a change of the frame of reference. This would allow for considering the problem in the frame comoving with the detector. It would also involve decomposing the field in a different basis of modes. Such an analysis is, in principle, very much desirable but also much more challenging technically. On the other hand such an approach would make possible the investigation of nonclassicality beyond its operational formulation. This could help to address more general issues, one of them being the robustness of the classicality of a state with respect to the change of the mode decomposition.

Other open questions emerging from our analysis refer to the presence of the critical velocity and the resonance effect, and the appearance of the asymptote for a non-inertial detector. We suspect that the former phenomenon is due the specific method used, and the latter one could originate from the detector being affected by the Unruh particles. Further studies on this topic are required to better understand these phenomena. 

Also, to obtain more insight, different scenarious could be studied, such as considering the detector in free space. Furthermore, different approaches to nonclassicality could be investigated \cite{Richter,Rabl}.


\section{Acknowledgements}
The authors would like acknowledge financial support
from the National Science Center, Sonata BIS Grant No.
DEC-2012/07/E/ST2/01402.

\appendix
\section{Time evolution operator}
Here we show the steps allowing us to obtain the time evolution operator describing the forced harmonic oscillator \cite{fho}. 
Consider a harmonic oscillator undergoing the evolution described by the time-dependent Hamiltonian of the form $\hat{H}=\hat{H}_0+\hat{V}(\tau)$. 
We choose to work in the interaction picture with respect to the Hamiltonian $\hat{H}_0$, and therefore introduce:
\begin{align}
&\hat{V_I}(\tau)=e^{i\hat{H}_0\tau}\hat{V}(\tau)e^{-i\hat{H}_0\tau},\nonumber\\ 
&|\psi(\tau)\rangle_I=e^{i\hat{H}_0\tau}|\psi(\tau)\rangle,
\end{align}
and $i\frac{d}{d\tau}|\psi(\tau)\rangle_I=\hat{V}_I(\tau)|\psi(\tau)\rangle_I$. The time evolution operator is transformed according to the equation: 
\begin{align}
\hat{U}_I(\tau,\tau_0)&=e^{i\hat{H}_0\tau}\hat{U}(\tau,\tau_0)e^{-i\hat{H}_0\tau},
\end{align}
and satisfies the following relation: $i\frac{d}{d\tau}\hat{U}_I(\tau,\tau_0)=\hat{V}_I(\tau)\hat{U}_I(\tau,\tau_0)$, a formal solution to which gives us:
\begin{align}
\hat{U}(\tau,\tau_0)&=\mathcal{T}\bigg\{\exp\big[-i\int_{\tau_0}^\tau\hat{V}_I(\tau')\text{d}\tau'\big] \bigg\}.
\end{align}
In the above formula $\mathcal{T}$ stands for the time-ordering operator. If we assume that the operator $\hat{V}(\tau)$ satisfies the relation:
 \begin{align}
 [\hat{V}_I(\tau),[\hat{V}_I(\tau'),\hat{V}_I(\tau'')]]=0,\label{App-rel}
 \end{align}
 then the time evolution operator can be simplified to:
\begin{align}
\hat{U}_I(\tau,\tau_0)&=\exp\bigg[-i\int_{\tau_0}^\tau\hat{V}_I(\tau')\text{d}\tau'\nonumber\\
&-\frac{1}{2}\int_{\tau_0}^\tau \text{d}\tau'\int_{\tau_0}^{\tau'}\text{d}\tau''[\hat{V}_I(\tau'),\hat{V}_I(\tau'')]\bigg].
\end{align}
The explicit form of the Hamiltonian $\hat{H}_0$ and the operator $\hat{V}(\tau)$ taken from the model of the harmonic oscillator is: 
$\hat{H}_0=\omega \hat{a}^\dagger \hat{a}$ and $\hat{V}(\tau)=f(\tau)(\hat{a}+\hat{a}^\dagger)$. Therefore we can calculate the commutator: 
\begin{align}
[\hat{V}_I(\tau'),\hat{V}_I(\tau'')]=-2if(\tau')f(\tau'')\sin(\omega(\tau'-\tau'')),
\end{align}
and confirm that $\hat{V}_I(\tau)$ satisfies the relation \eqref{App-rel}. Based on this observation we can calculate the time evolution operator to obtain: 
\begin{align}
\hat{U}_I(\tau,\tau_0)=e^{i\beta(\tau,\tau_0)}\hat{D}(\zeta(\tau,\tau_0)).
\end{align} 
In the above $\hat{D}(x)=\exp\{x \hat{a}^\dagger-x^* \hat{a}\}$ denotes the displacement operator and we have introduced:
\begin{align}
\zeta(\tau,\tau_0)&=-i\int_{\tau_0}^\tau f(\tau')e^{i\omega \tau'}\text{d}\tau', \\
\beta(\tau,\tau_0)&=\int_{\tau_0}^\tau \text{d}\tau'\int \text{d}\tau'' f(\tau')f(\tau'')\sin(\omega(\tau'-\tau'')).
\end{align}
The above formulae can be used for evaluating the time evolution operator $\hat{U}_{k,\pm}(\tau)$ discussed in the core body of the manuscript.\\

\section{Coherent and thermal states}
Recall that the $P$-representation for coherent and thermal states take the following form (here we consider one mode only and omit the label $k_0$):
\begin{align}
&P_{\text{coh}}(\alpha)=\delta(\alpha-\alpha_0),\\
&P_{\text{th}}(\alpha)=\frac{1}{\pi\bar{n}}\exp(-|\alpha|^2/\bar{n}),
\end{align}
where $\bar{n}$ is the average number of excitations. One can insert these formulae into the first line of equation (\ref{witness-moving}) and evaluate the absolute value of the witness function. For the coherent state the corresponding absolute value of the witness function is equal to one, whereas for the thermal state we obtain:
\begin{align}
\left|W_{\text{th}}(\tau)\right|&=\left|\int \text{d}^2\alpha P_{{\text{th}}}(\alpha)\exp\{4i\text{Im}(\alpha^*\chi(\tau))\}\right|\nonumber\\
&=\left|\int \text{d}^2\alpha \exp\{-\frac{1}{\bar{n}}|\alpha|^2-4i\text{Im}(\alpha^*\chi(\tau))\}\right|\nonumber\\
&=\exp\left\{-4\bar{n}|\chi(\tau)|^2\right\} \leq 1.
\end{align}
In both cases we obtain a function which does not exceed unity.
\section{Resonant behavior}
Here we evaluate the expression for $\chi_{k_0}(\tau)$ that appears in Eq.~\eqref{chi_v}. Let us repeat the underlying integral:
\begin{align}
\chi_{k_0}(\tau)=\frac{-i\lambda}{\sqrt{k_0\pi}}\int_0^\tau &\text{d}\tau'\sin\left(k_0\pi\left(\frac{1}{2}+\frac{x_0}{L}\right)+\frac{k_0\pi v\,\tau'}{L\sqrt{1-v^2}}\right),
\end{align}
and the notation $\omega_L=\frac{k_0\pi v}{L\sqrt{1-v^2}}$. For further convenience we introduce $\varphi=k_0\pi\left(\frac{1}{2}+\frac{x_0}{L}\right)$ and continue the calculation to obtain: 
\begin{widetext}
\begin{align}
\chi_{k_0}(\tau)&=\frac{\lambda  \left(e^{i \tau  \omega _{k_0}} \left(\omega _{k_0} \sin \left(\tau  \omega _L+\varphi \right)+i \omega _L \cos \left(\tau  \omega
   _L+\varphi \right)\right)-\omega _{k_0} \sin (\varphi )-i \omega _L \cos (\varphi )\right)}{\sqrt{\pi } \sqrt{k} \left(\omega _L^2-\omega
   _{k_0}^2\right)}.
   \end{align}
\end{widetext}
This expression reveals a resonant behavior with a peak at $\omega_{k_0}=\omega_L$, namely if we take the limit $\omega_{k_0}\to\omega_L$, we obtain a finite $|\chi_{k_0}(\tau)|$ whose envelope increases linearly with $\tau$. It allows us to identify the critical velocity given in Eq.~\eqref{v-c}.

\end{document}